\documentclass[aps,prbbib]{revtex4}
\usepackage{isolatin1}
\usepackage[T1]{fontenc}
\usepackage{times}

\usepackage{graphicx}
\usepackage[dvips]{color}
\begin{document}
\title{Extracting Structural Information of a Heteropolymer from Force-Extension Curves}
\author{Elena Jarkova$^{1}$}
\author{Nam-Kyung Lee$^{1,2}\footnote{e-mail: lee@sejong.ac.kr}$}
\author{Sergei Obukhov$^{1,3}$}
\affiliation{Institut Charles Sadron , 67083 Strasbourg Cedex, France.}
\affiliation{Department of Physics, Sejong University, Seoul 143-743, South Korea}
\affiliation{Department of Physics, University of Florida, Gainesville, FL32611
}
\date{\today}

\begin{abstract}
We present a theory for the reverse analysis on the sequence
information of a single H/P two-letter random hetero-polymer (RHP)
from its force-extension ($f-z$) curves during quasi static
stretching. Upon stretching of a self-assembled RHP, it
undergoes several structural transitions. The typical elastic
response of a hetero-polymeric globule is a set of overlapping
saw-tooth patterns. With consideration of the height and the
position of the overlapping saw-tooth shape, we analyze the
possibility of extracting the binding energies of the internal
domains and the corresponding block sizes of the contributing
conformations.
\end{abstract}

\maketitle
\section{Introduction}
Within a last decade, a series of remarkable force-extension
experiments was performed using Atomic Force Microscopy (AFM).
These experiments show that the elastic response of a single
molecule is clearly related to the internal structure of the
molecule. Force-extension profiles of single molecules  such as DNA,
RNA, synthetic polyelectrolytes, giant protein titin and chromatin
fibers
\cite{rief97,kellermayer97,senden:98.1,li:98.1,courvoisier:98.1,li00,liphardt01,gao03,bro02}
show characteristic saw-tooth patterns, which are interpreted as successive unfolding of
internal domains. This is in agreement with theoretical
studies\cite{lee02c,lee02a,lee02b,tamashiro00,geissler02} and computer simulations
\cite{klimov99,paci00,lu00}  predicting that step-wise unfolding
pattern can be seen from the unfolding of pearl necklace of
polyelectrolytes in a poor solvents\cite{tamashiro00,vilgis00} and protein
models \cite{klimov99}.

In some polymer systems
(specially, biopolymers and proteins), the intra-chain
self-assembly produces secondary or tertiary structures  and the
elastic response reflects this structural hierarchy\cite{gao03}.
The AFM experiments show that a series of partial unfoldings of those collapsed structure occurs by  applying an external force.
When the elastic energy gain is comparable with increase of the potential energy, the extension increases abruptly by $\delta z$.
The resulting force-extension profile is rich and reflects the domain size responding to the
applied force.  Information on the sequence of the linear structure reveals on the force-extension curve.
In this sense it is interesting to trace back
the particular sequence structure of a given chain from the measured elastic response.

In our previous study\cite{lee02a,lee02b}, we minimized the free
energy at the given force (which mimics a constant force measurement
experiments). The obtained minimum corresponds to the ground state
or to the metastable states.  At several characteristic values of
force, segments of linear chain in the collapsed phase unfold in
the pattern of ``plateaus'' in $f-z$ curve. However, these
``plateaus'' often correspond to the multiple conformational
transitions going through different extensions $z$ if domains have
similar binding energies. Therefore sequence information is partly
washed away under the constant force measurements.

Another experimentally common, yet theoretically more challenging,
set up of AFM measurement is performed by imposing the distance
and measuring the restoring force. Typically, the force-extension
profile has a saw-tooth pattern.  Each time an internal domain is
pulled out, the contact with cantilever becomes loose resulting in
a big drop of the measured force. Hence, this sequence information
is more directly accessible by force-extension measurement when
the distance is imposed. Then an arising question is, if it is
possible to recover the information about the sequence of polymer
from force-extension profiles.
 For this purpose, we present theoretical frame work how to
"read" the sequence information from the  elastic response.
We demonstrate the mapping of the
 force-extension profiles to the sequence information
under the controlled displacement. We show that it is feasible to
extract the composition of block sizes to some extent while the
order of arrangement of those blocks still remains to be answered.

\section{\textbf{General model}}
\label{sec:method}
We consider a polymer
chain of $N$ monomers, one end of which is fixed at a reference point (i.e. $z=0$) and the other end is brought to the distance $z$  from the reference point
. The sequence consists of $n_h$ of
hydrophobic (h) blocks and $n_p$ of hydrophilic (p) blocks in an
alternating order ($n_h=n_p$). The size of i-th
hydrophobic (hydrophilic) segment is
 $N_i^{h}$ ($N_i^p$)
and the sequence of the whole chain can be represented by a series
of h- and p- blocks of sizes: $\{ N_i^p,N_i^h\}$.

We assume that hydrophobic segments $N_i^h$ have a tendency to
collapse into a compact globule of radius $a_i$
and these unit globules are not further stretchable. These single
block globules can merge  together into a larger globule
leaving the connecting hydrophilic segments as loops on the
surface of the large globule.

The optimal conformation of the chain is obtained from the minimum
free energy  under the given extension $z$. The  free energy
consists of the two main contributions, the interaction energy of
the collapsed h-blocks (globules) and the elastic part of the
p-blocks (strings).
For simplicity, here we assume that p-block strings have the
elastic properties of ideal Gaussian chain (later we discuss more
realistic Langevin chain model).
For the chain of length $N_i^p$ and size $z_i$, the elastic  energy
is $F_{elastic}=z_i^2/N_i^p$. The elastic part of the free energy
comes from the released hydrophilic segment connecting two nearest
globules. Loops (hydrophilic segments whose both ends are attached
to aggregated globule) do not contribute to the total elastic
energy.

\begin{figure}[tbp]
\includegraphics[width=6.5cm]{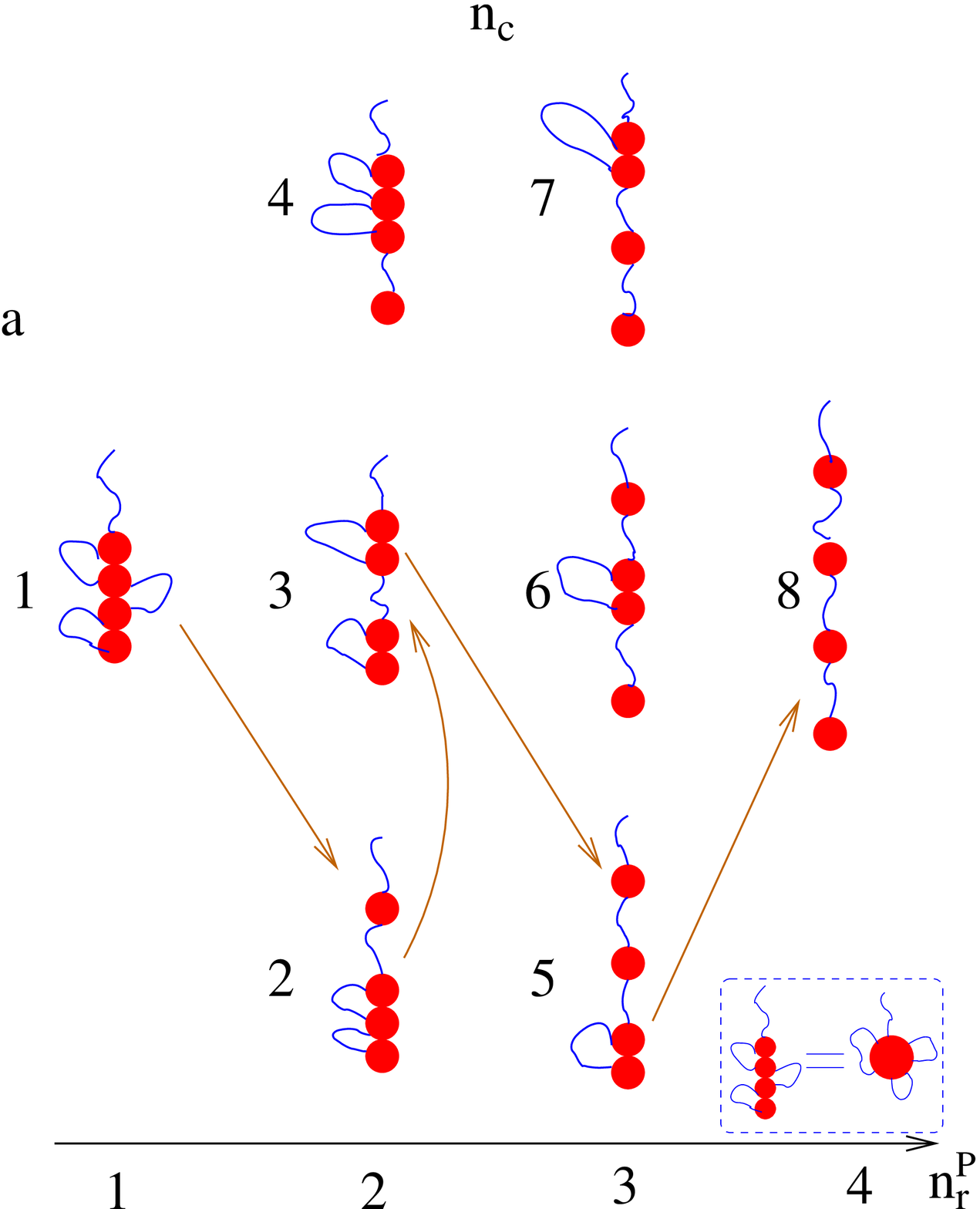}\hspace{1.5cm}
\includegraphics[width=9.cm]{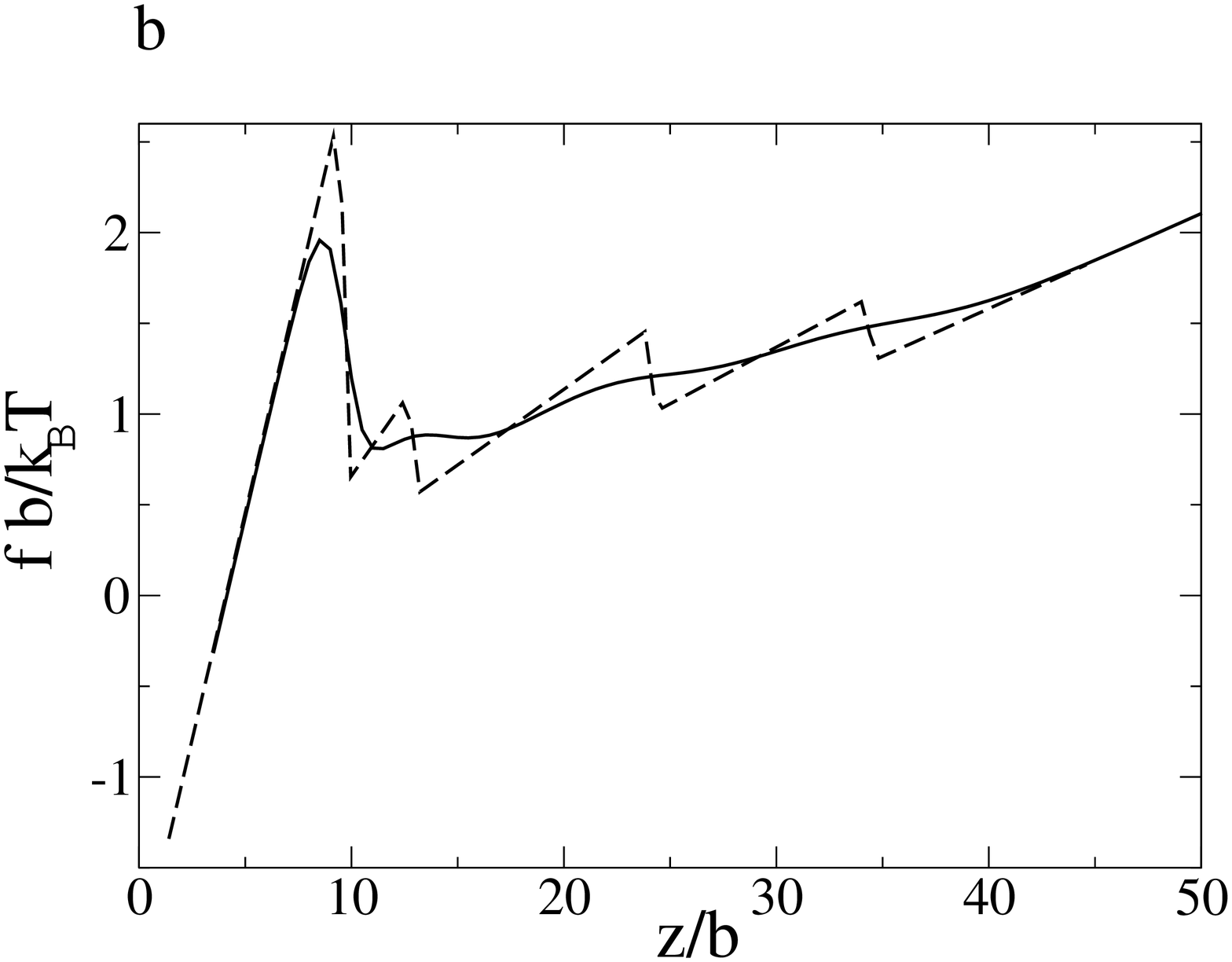}
\caption{(a) Schematic picture of all possible conformations for a
heteropolymer consisting of four h- and p-blocks
of sequence 4p-4h-8p-6h-20p-6h-6p-6h, classified according to
the number of the released p-blocks  ($n_r^p$).
(b) The force-extension curves for the same sequence.
The dashed-line is the force-extension curve
following the minimal energy conformations according to Eq.(\ref{eq:freeenergy}) with $\gamma =9 k_BT/b^2$ ($\tau =3$) and the solid-line is
the statistically averaged force-extension.
}
\label{fig:struc}
\end{figure}

Initially the chain is fixed at the minimal extension $z_0$
($z_0 \ll Nb$),
 all h-blocks belong to a large collapsed globule
and only one end p-block is outside of this globule.
As imposed distance $z$ varies, the chain adapts its conformation
in order to minimize the total free energy.
Each conformation can be characterized by the numbers of the
released p-blocks and the position of the released p-blocks as
illustrated in Fig.\ref{fig:struc}(a). The free energy of the
conformation of which $q$-th p-block is released is written as:
\begin{eqnarray}
\frac{E}{k_{B} T}=\frac{\gamma b^2}{k_{B} T}\left(S_1^q+
S_{q+1}^{n_h}\right) +\frac{\left(z-2 b\left(a_1^q +
a_{q+1}^{n_h}\right)\right)^2}{(N^p_1+N^p_q)b^2}.
\label{eq:freeenergy}
\end{eqnarray}
where $S_k^m$ and $a_k^m $ denote the surface area and the radius of the globule consisting of $k,k+1,\dots ,m$-th h-blocks, respectively 
and $\gamma=k_B T \tau^2/b^2$ is a surface tension with $\tau$ being reduced temperature $\tau = |T-\theta|/\theta$.
Similar equations can be written for the conformations with the
arbitrary number of the h-blocks. If there are $n_{p}$ p-blocks,
there are $n_p-1$ conformations of which one of the internal p-block
is released.
The number of conformations where $m$ out of $n_p$ p-blocks are
released is $_{n_p}C_{m} = \frac{m!}{(m-n_p)!n_p!}$. The total
number of conformation is $\Omega= 2^{n_p-1}$. In
Fig.\ref{fig:struc}(a), we show all $2^{4-1}=8$ possible conformations
of a heteropolymer consisting of 4 p-blocks and h-blocks.  The
conformations listed along the vertical lines have the same number
of released p-blocks ($n^p_r$) but different grouping of
h-blocks. 

In general, the free energy of each conformation is
slightly different from each other.
For any given extension $z$, there are several local energy minima
with similar free energy $E_r$. These  conformations  contribute
to the thermodynamic properties of the force-extension relation
with statistical weight of $\exp(-E_r/k_BT)$. In order to plot the
force-extension curve, all possible conformations at given $z$
must be taken into account with this statistical weight. The
statistical sum $G(z)$ of all possible conformations at the displacement
$z$ is:
\begin{eqnarray}
G(z)=\sum_{r} \exp \left(-\frac{E_r(z)}{k_{\rm B} T}\right).
\label{eq:gz}
\end{eqnarray}
 The restoring force acting on the polymer chain is:
\begin{eqnarray}
f=- k_BT \frac{\partial \ln G(z) }{\partial z}. \label{eq:f}
\end{eqnarray}


In Fig.\ref{fig:struc}(b) we show a force-extension curve
calculated for a randomly chosen sequence
4p-4h-8p-6h-20p-6h-6p-6h. For convenience, we choose
$S_k^{m}=b^2(\sum_{i=k}^{m}N^h_i/\tau)^{2/3}$ and
$a_k^{m}=b(\sum_{i=k}^{m}N^h_i/\tau)^{1/3}$\cite{note}. The dashed-line represents the force obtained from minimal energy
conformation for each extension $z$. If fluctuation is negligible,
the expected force-extension curve is a sharp saw-tooth
pattern shown as dashed-line in Fig.\ref{fig:struc}(b). Each
transition from one conformation to another is
captured as a "drop" of a restoring force, which indicates the minimum energy
conformation switches into the different conformation.
The force increases with the extension until the next "drop".
The force between the "drops" is proportional to  $%
\sim{z/N_{p}}$ ($N_p$ is a sum of free p-blocks).
The longer the chain is, the easier to stretch it.

The solid line is the force-extension curve obtained from
Eq.\ref{eq:gz} and Eq.\ref{eq:f} where all local energy minima
conformations are also taken into account with proper statistical
weight.  The unfolding of the large globule
follows the path illustrated in the Fig.\ref{fig:struc}(a).
Release of each unit globule leads to a jump.  The height of each
jump becomes smaller as the overall globule size becomes smaller
so that surface energy difference before and after the release
becomes smaller. We  note that one of the transitions is between
conformations with the same number of released p-blocks. (2nd jump
from conformation 2 to the conformation 3.)

\begin{figure}[tbp]
\includegraphics[width=7cm]{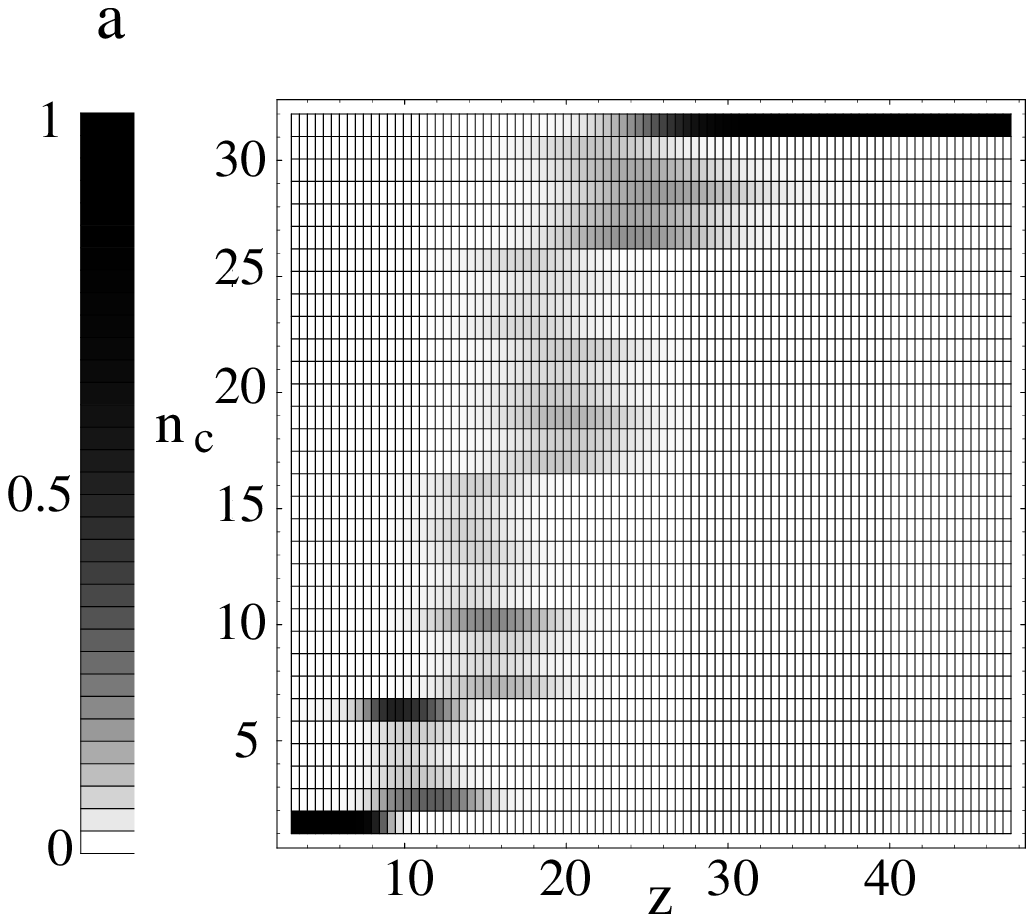}\hspace{1.cm}
\includegraphics[width=9cm]{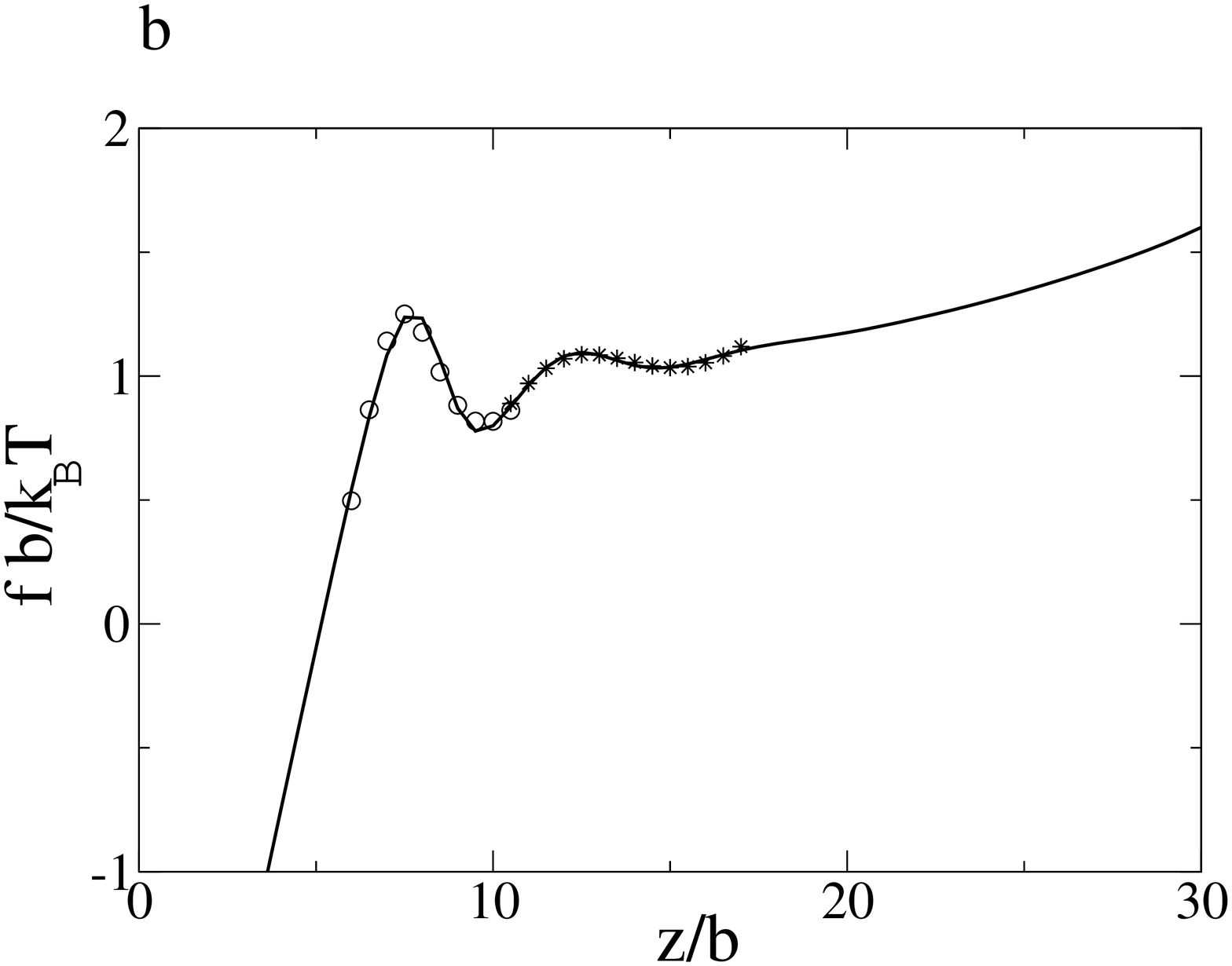}
\caption{The probability distribution of a sequence
3p-6h-5p-5h-3p-3h-2p-2h-3p-10h-3p-3h in the space of all possible
conformations with $\gamma=4 k_{\rm B}T/b^2$ ($\tau =2$).
x-axes is the given extension $z$,
y-axes is the index of the each conformations defined in the similar way illustrated in Fig.\ref{fig:struc}.
The grey scale bar in the left hand side shows the probability scale.
(b) The corresponding force-extension curve. The symbols ($\circ$
and $+$) represent the fitting results using Eq.\ref{eq:fit}.
}
\label{fig:force_path}
\end{figure}

Another example of force extension curve for a different randomly
chosen sequence(6 p-blocks and 6 h-blocks of different lengths) is shown on
 Fig.\ref{fig:force_path}.
Here the probability to be at each conformation $i$ (given by
Eq.(\ref{eq:gz})) is shown in the space of all possible $2^5$
conformations, (see Fig.\ref{fig:force_path}(a)). The dark region
indicates the favorable conformations under given constraint
(fixed $z$). In  some range of $z$, there are several
conformations with similar statistical weight. The transition from
one group of conformations to another on the
Fig.\ref{fig:force_path}(a)  near $z=20$ does not result in any
noticeable feature on  (Fig.\ref{fig:force_path}(b)). There are
visible only few first jumps corresponding to the conformational
transitions.

Why only two or three transitions are visible on the force-extension curve?
At the jump, dominant conformation shifts from one to the other.
If there is a clear favorite conformation, the transition is
sharp. Otherwise, if several conformations contribute with similar
weights, transitions are not expected to be captured as a clear
saw-tooth shape in the quasistatic measurement and the fluctuation
around average force is large.
Around each transition there is a
region of strong  fluctuations, $\delta z_n$ where difference in energy of competing conformations is smaller than $k_{\rm B}T$. For
the n-th transition this region is about $\delta z_n = k_{\rm B}T
z_n^*/\varepsilon_n$. Here $z^*_n$ is n-th transition point and
$\varepsilon_n$ is the binding energy related to this transition.
The size of fluctuation region is typically growing with $n$
because $z^*_{n}\propto n$. We should note that the binding energy
$\varepsilon_n$ can not be much larger than $k_{\rm B}T$,
otherwise it is difficult to perform quasistatic experiment. It
means that  after several transitions $n\approx
\varepsilon_n/k_{\rm B}T$ their fluctuation regions should
overlap: $ \delta z \approx z^*_{n} - z^*_{n+1} $ and the typical
zigzag pattern of each transition starts overlapping with that of
neighboring transitions. Fig.\ref{fig:force_path} (a) demonstrate
such smooth $f-z$ curve after a few initial jumps. At large
extension, when all loops are pulled out, force increases
monotonically with extension.

In realistic experimental situations, one chain end is pulled with
a small but finite speed. The free energy difference $\delta E^b$
between the dark and bright conformation gives typical relaxation
time $\sim e^{\delta {E^b/{k_B T}}}$ for the transition between two likely
conformations. Depending on the pulling speed, certain energy
barrier conformations are overcome but some of them are not.
Conformations separated by the large energy barrier do not
contribute the $f-z$ when the pulling speed is faster than the
chain relaxation time. Thus, the accessible conformation can be
controlled by pulling rate and this allows extracting more
detailed information about the structure of polymer. We shall
address this question in the future publication.

\section{Reading the sequence information from f-z curve}
\begin{figure}[tbp]
\includegraphics[width=12cm]{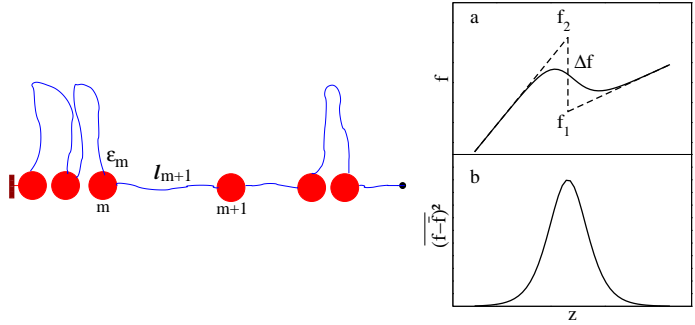}
\caption{The simple model of copolymer. (a) Typical f-z curve and (b) fluctuations in $f$ around a transition point.}
\label{fig:simple}
\end{figure}

{\bf Simple Model}
In the following, we show how to extract the chain sequence
information 
from  the force-extension curve.
In order to do so, we further simplify the conformational space.
As illustrated in Fig.\ref{fig:simple}, we
assume that globules are arranged in 1-d and interact only with neighboring
globules. We denote $\varepsilon_{m}$ as
the interaction energy between $m$-th and $m+1$-th h-block globules.
The transition in conformations is related only to the releasing of a unit globule-loop pair from a larger globule.
We will show that the interaction energy, $\varepsilon_m$, can be extracted
from the analysis of the force extension curve (see Fig.\ref{fig:struc}).
More realistic assumption
would be that all aggregated globules $m-2, m-1, m$  interact with $m+1$-th
globule.
In this case the energy $\varepsilon_m$ depends on the
arrangement of globules.

In 1-d model, each conformation is
completely characterized by two sets of  variables: $\{\varepsilon _{m}\}$ and $\{l_{m}\}$, where $\varepsilon_{m}=\gamma \left[S_{1}^{n_h}-(S_1^m+S_{m+1}^{n_h})\right]$, $S_n^m$ is a surface area of globule consisting of $N_n^h, N_{n+1}^h,..., N_ m^{h}$
h-blocks
and $l_{m}$ is the length of the $m-$ th p- block $l_{m}=N_{m}^{p}b$.
 In the absence of an external
force, all h-globules are attached and aligned in one line.
With  the increase of the applied distance $z$, the contacts
between h-globules break one after another. In the force-extension
curve, these events are represented as "drops" in force. The
phenomenological knowledge of  $z$-coordinate of the jump (denoted as $z^*$ below) and its
magnitude $\Delta f$ allows to  determine $\varepsilon _{m}$ and
$l_{m}$, uniquely. At the conformational transition of releasing $m+1-$th
loop,where $z=z^*$, the energies
of two conformations should be equal: $E_{m} = E_{m+1}$. This
leads to the following relation in $k_BT$ units,
\begin{equation}
\frac{{z}^{2}}{L_m b}=\frac{{z}^{2}%
}{(L_m+l_{m+1})b}+\varepsilon _{m+1}  \label{1} \label{eq:reading1}
\end{equation}
where $L_m$ is the total linear length of the chain before the
transition.
In this relation we assume that the p-block segments are much
longer than the size of the collapsed h-blocks.
Otherwise, the size of the hydrophobic globules becomes relevant as an offset
of elastic energy of the chain.
Then Eq.\ref{eq:reading1} reads
\begin{equation}
\frac{(z- 2a_{1}^{n_h})^{2}}{L_{m}}=\frac{(z-2 (a_{1}^m+a_{m+1}^{n_h}))^{2}}{L_{m}+l_{m+1}}+\varepsilon_{m+1}
\end{equation}
One can relate the height of this jump $\Delta
f=f_{m}-f_{m+1}=\partial E_{m}/\partial z-\partial
E_{m+1}/\partial z= [2z/L_m -2z/\left(L_m+l_{m+1}\right)] (k_BT/b) $ with $\varepsilon _{m+1}$:
\begin{equation}
\varepsilon _{m+1}=\frac{\Delta fz}{2} \label{eq:fit1}
\end{equation}
Similarly, the length $l_{m+1}$ can be extracted from the slope
difference before and after the jump
$1/f_{m+1}-1/f_{m}=l_{m+1}/2z$:
\begin{equation}
l_{m+1}=\frac{2\Delta fz}{f_{m+1}f_{m}}\frac{k_BT}{b}
\label{eq:fit2}
\end{equation}

We notice that all inclined parts of the curve, if continued, have
zero intercept. The order of releasing is determined by either the
minimal interaction energy
of h-globules  $%
\varepsilon _{m}$ among all remaining $\varepsilon _{k}$\ or the
maximal length of the p-segment $l_{k}$. If all blocks are of
similar size with small
variations $\delta \varepsilon _{k}$, $\delta l_{k}$ around the average values $%
\varepsilon $ and $l$, 
then $L_m \approx z$
and from
Eq.(\ref{1}) we can get the condition of releasing the next
segment $k$ as determined by the largest relative variation $\max\{\max (\frac{%
-\delta \varepsilon _{k}}{\varepsilon }),\max (\frac{\delta
l_{k}}{l})\}$.

{\bf  Reading thermally averaged f-z curves }
In the vicinity of the m-th transition point $z_m^{\ast}$ (below we simplify as $z^{\ast}$) where (m+1)-th loop is released, the difference between the
energies of two states can be small and comparable to $k_{\rm B}T$. Because
of thermal fluctuations, the actual force-extension curve can be very noisy.
If these fluctuations are properly averaged, 
the edge of sharp saw-tooth is rounded.
(see Fig.\ref{fig:simple}). We will show below, that parameters
$\varepsilon_{m}$ and $l_{m}$ can be extracted from rounded curve
too.

In the absence of thermal fluctuations we describe the jump in a
force-extension curve in Fig.\ref{fig:simple} using well-known
step-function ($\theta$-function): $f\left( z\right) =f_{m}\left(
z\right) \theta \left( z^{\ast }-z\right) +f_{m+1}\left( z\right)
\theta \left( z-z^{\ast }\right) $, where $f_{m}\left( z\right) $
, $f_{m+1}\left( z\right) $ are force-extension curves before and
after the jump ($f_{m}\left( z\right) =\partial E_{m}\left(
z\right) /\partial z$). In order to include the rounding effect of
thermal fluctuations we will replace $\theta$-function in this
equation by thermally averaged function $\bar{\theta }$:
\begin{equation}
\bar{\theta}\left( z^{\ast}-z\right)
=\frac{e^{-E_{m}\left( z\right)
/k_{\rm B}T}}{e^{-E_{m}\left( z\right) /k_{\rm B}T}+e^{-E_{m+1}\left( z\right) /k_{\rm B}T}}=\frac{1%
}{1+e^{-\left( E_{m+1}\left( z\right) -E_{m}\left( z\right)
\right) /k_{\rm B}T}}
\end{equation}
At the transition point, where $E_{m+1}\left(z\right)
-E_{m}\left(z\right) =0$, thermally averaged function is
$\bar{\theta }\left( z^{\ast
}-z\right) =1/2$.
In the vicinity of the transition we can interpolate difference $%
E_{m+1}\left( z\right) -E_{m}\left( z\right) $ as $2\varepsilon
_{m+1}(z^{\ast }-z)/z^{\ast }$. Finally we obtain:
\begin{equation}
\bar{\theta }\left( z^{\ast }-z\right)
=\frac{1}{1+e^{-\frac{z^{\ast }-z}{z^{\ast }}\frac{2\varepsilon
_{m+1}}{k_{\rm B}T}}}
\label{eq:avetheta}
\end{equation}
and the fitting function for a transition is:
\begin{equation}
\overline{f\left( z\right)} =f_{m}\left( z\right) \bar{\theta }\left(
z^{\ast }-z\right) +f_{m+1}\left( z_m\right) \bar{\theta
}\left( z-z^{\ast }\right). \label{eq:fit}
 \end{equation}
There are three independent variables controlling the shape of a single saw-tooth jump (see Fig.\ref{fig:simple}): slopes before and after transition and the location of the transition $z^*$, the same number of independent
variables is needed for  fitting of the thermally averaged curves. Each
additional transition requires two additional
variables for its description: $z$-coordinate of transition and
the slope after the transition, which can be related to $l_m$ and
 $\varepsilon_m$ through Eqs.\ref{eq:fit1} and \ref{eq:fit2}.

If two or more transitions are close to each other, then it might
be difficult to determine the slope of the force-extension curve
in the regions between these transitions, especially with presence
of noise. Here we present fitting functions for two overlapping
transitions, the further generalization for multiple transitions
is obvious. The combined fitting functions for two transitions can
be symbolically written with the use of $\theta$ functions as:
$\overline{f(z)} =f_{m}\left( z\right) \bar{\theta }\left(
z_m^{\ast }-z\right) +f_{m+1}\left( z\right) \overline{\theta
\left( z-z_m^{\ast }\right)\theta \left(
z_{m+1}^{\ast}-z\right)}+f_{m+2}\left( z\right) \bar{\theta
}\left( z-z_{m+1}^{\ast }\right)$.

Here the averaged product of two $\bar{\theta}$ functions
represents
\begin{equation}
\overline{\theta\left( z-z_m^{\ast }\right) \theta
\left( z_{m+1}^{\ast}-z\right)}  = \frac{e^{-E_{m+1}}/k_{\rm
B}T}{e^{-E_{m}/k_{\rm B}T } + e^{-E_{m+1} /k_{\rm
B}T}+e^{-E_{m+2}/k_{\rm B}T }}.
\label{eq:prodtheta}
\end{equation}
It should be noted that the locations of the released loops of
both transitions are not necessarily next to each other along the
chain. After global optimization over fitting parameters, we
produce the best estimate for this circumstance.  If transitions
are too close to each other ($(z_m^{\ast }-z_{m+1}^{\ast })/{z} <
k_{\rm B}T/2\varepsilon_m$), the fitting curve gives better estimate
of
the sum of energies $\varepsilon _{m+1}+$ $\varepsilon _{m+2}$ and lengths $%
l_{m+1}+l_{m+2}$, but not estimate of these quantities by
themselves.

The symbols in Fig.\ref{fig:force_path}(b) ($\circ$ and $+$)
represent the fitting results of the function, Eq.\ref{eq:fit}.
When the first h-block is released the unknown parameters are
 the transitional point $z^*$ and
 the lengths of the released p-blocks before and after the transition:
 $l_1$ and $l_1+l_2$. Notice, that in the case when the total size of globules on the string before and after event, $\alpha_m$ and $\alpha_{m+1}$, are not small one should consider them as  additional fitting parameters, so that force has a form $f_m=\frac{2\left(z-\alpha_m\right)k_{\rm B}T}{L_m b}$.
The best fit is  obtained with parameters $z^*\sim 8.3 b $, $l_1\sim
3.0 b$ and $l_1+l_2\sim 8.0 b$, $\alpha_1\sim 5.2 b$ and
$\alpha_2\sim 7.2 b$. When the second h-block is released ($+$)
 $z^*$, $l_1+l_2$ and $l_1+l_2+l_3$ are unknown.
 From the second event, we obtain,
$z^*\sim 13.4 b$, $l_1+l_2\sim 9.1 b$ and $l_1+l_2+l_3\sim 11.0 b$,
$\alpha_2\sim 6.3 b$ and $\alpha_3\sim 10.3 b$.

After all we have $l_1 = 3.0 b$, $l_2 = 5.0 b$ and $l_3 = 1.9 b$,
which are in agreement with the exact values for p-blocks $3, 5,
3$ accordingly. The estimated  interaction energy difference
before and after event from Eq.\ref{eq:reading1},
$\varepsilon_{m}=\left(f_m\left(z^*-\alpha_m\right)-f_{m+1}\left(z^*-\alpha_{m+1}\right)\right)/2$,
are $\Delta\varepsilon_1/k_{\rm B}T\sim 3.1$ and $\Delta\varepsilon_2/k_{\rm B}T\sim 4.7$. The estimated interaction strengths are in agreement with
the calculated values $\varepsilon_1/k_{\rm B}T=\gamma/k_{\rm B}T \left((S_1^5)+(S_6^6)-(S_1^6)\right)=4.9$ and
$\varepsilon_2/k_{\rm B}T=\gamma/k_{\rm B}T \left((S_1^4)+(S_5^5)-(S_1^5)\right)=4.3$.

{\bf Matching the noise pattern} The above fitting was done to the
thermodynamically averaged transition curve. In practice this
curve  can be quite noisy especially in the transition region,
because system fluctuates between two different configurations
with similar energies and time averaging could be costly. It makes
sense to measure the noise directly as a function of
extension $z$ and try to extract structural information from it.
Calculating the average mean-square magnitude of thermal noise we
get:
\begin{equation}
\overline{(f(z)-\overline{f(z)})^2}=\bar {\theta} (z^*-z)\cdot
\bar{\theta} (z-z^*) \frac{{4{\varepsilon_{m+1}}^2}}{z^{\ast {2}}}
\end{equation}
This function is the product of two thermally averaged $\bar{\theta}$ function 
defined in Eq.\ref{eq:avetheta}  and is sharply peaked as is shown on the second inset of Fig.\ref{fig:simple}(b).

{\bf Langevin chain} For the practical application, we consider
the Langevin chain (with fixed bond length) for which the chain
extension is given by the following Langevin equation.
\begin{equation}
z/L= \left [\coth\left(\frac{fb}{k_{\rm B}T}\right) - \frac{k_{\rm B}T}{fb}\right ]
\end{equation}
In the limit of  strong stretching, this equation can be simplified to $fb/k_BT \simeq 1/(1-z/L)$ and for weak
stretching limit, it reproduces the linear response behavior $fb/k_BT \simeq  z/L$. We can assume, that before
and after transition point, the $f-z$ curve is described by strong  and weak
stretching behavior, respectively.  Than instead of  Eq.\ref{eq:fit2}  we have:
\begin{equation}
l_{m+1} = z \left [\frac{1}{1- k_{\rm B}T/f_m b} - \frac{3}{f_{m+1}b/k_{\rm B}T} \right].
\end{equation}
This reading method can be applied to the experimental curve of the protein
domain unfolding where each saw-tooth (jump) corresponds to the unraveling of
a single domain. We do not try to fit the detail shape of the curve which
often treated as worm-like-chain model. We note that the position of peaks
and the depth of the jump can be directly mapped into our
1-dim globule-string model.   We may map the number of monomers in the each domain
into the connecting p-block size in our model because after the unfolding of
each domain, the extension increases by the length corresponding domain size. The binding energy of the each domain is now the interaction energy between two h-globules, i.e $\varepsilon_m$.


\section{Conclusions}
We demonstrated that some structural information of heteropolymers can be
extracted from the force-extension curves using the simple model.  In this work,
we assume that the process of pulling is so slow thus system is always in thermodynamic equilibrium.  This means all possible conformations can contribute to  the elastic responses with appropriate thermodynamic weight.
In the future publication we will report the effect of finite pulling rate
where accessible number of configurations is controlled by
the pulling rate.

\textit{Acknowledgment} 
Authors are grateful to Dr. A.Johner (ICS) for his great hospitality, patience and  helpful
discussions. S.O. acknowledges LEA support.


\begin{thebibliography}{10}
\bibitem{note}
The surface area and radius of globule is not well
defined if the number of consisting monomers are too few.  We assume that
the aggregate has a perfect spherical shape.
\end{thebibliography}
\end{document}